\documentclass[review]{elsarticle}

\usepackage{lineno,hyperref}
\modulolinenumbers[5]

\journal{Journal of \LaTeX\ Templates}

\begin{document}

\begin{frontmatter}

\title{Yu-Shiba-Rusinov bands in superconductors in contact with a magnetic insulator}


\author[addresswb]{Wolfgang Belzig}
\cortext[mycorrespondingauthor]{Corresponding author}
\ead{Wolfgang.Belzig@uni-konstanz.de}

\author[addressdb]{Detlef Beckmann}

\address[addresswb]{Department of Physics, University of Konstanz, D-78457 Konstanz, Germany}
\address[addressdb]{Institute of Nanotechnology, Karlsruhe Institute of Technology (KIT), P.O. Box 3640, D-72021 Karlsruhe, Germany}

\begin{abstract}
Superconductor-Ferromagnet (SF) heterostructures are of interest due
to numerous phenomena related to the spin-dependent interaction of
Cooper pairs with the magnetization. Here we address the effects of a
magnetic insulator on the density of states of a superconductor based
on a recently developed boundary condition for strongly spin-dependent
interfaces. We show that the boundary to a magnetic insulator has a
similar effect like the presence of magnetic impurities. In particular
we find that the impurity effects of strongly scattering localized
spins leading to the formation of Shiba bands can be mapped onto the
boundary problem. 
\end{abstract}


\end{frontmatter}


Over the last two decades a tremendous progress in creating and
controlling heterostructures consisting of superconductors and
ferromagnets have been achieved both on an experimental and a
theoretical
level. \cite{Ryazanov:2001ua,Kontos:2001tx,Beckmann:2004dc,Keizer:2006jw,Khaire:2010eaa,Robinson:2010gia}. The
progress has been reviewed in
\cite{Bergeret:2005vj,BUZDIN:2005ua}. More recently, the experimental
focus has shifted toward magnetic insulators, offering certain
advantages like the absence of low-energy electronic excitations
responsible for the loss of superconducting coherence, and efficient spin filtering \cite{moodera1988,li2013b,pal2014}.
In this way, a long-range spin transport was demonstrated \cite{Hubler:2012if,wolf2014c} or
extremely large thermoelectric transport at low temperatures
\cite{Kolenda:2016ee,kolenda2017}. Recently tunneling specroscopy was
reported \cite{Strambini:2017wn}. On the theoretical side, it has
been shown that the proximity to magnetic insulator leads to a
suppression of the critical temperature \cite{MILLIS:1988ib} using
spin-dependent boundary conditions for the quasiclassical Green
functions. On the basis of a diffusive description, the induced
exchange splitting was used to suggest an absolute spin-valve effect
\cite{HuertasHernando:2002gj}. The boundary conditions have been
developed further \cite{Cottet:2009hya,Machon:2015wm} into the
present most general form \cite{Eschrig:2015cea}.

\begin{figure}[tb]
  \centering
  \includegraphics[width=8cm]{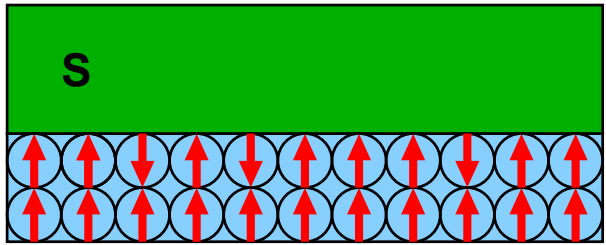}
  \caption{Sketch of a superconducting layer in contact with a magnetic
    insulator. Some of the surface spins might point in the opposite direction.}
  \label{fig:sketch}
\end{figure}

The quasiclassical problem of the density of states in a
superconductor close to a magnetic insulator as shown in
Fig.~\ref{fig:sketch} is readily formulated in
terms of the quantum circuit theory. We define the spin-dependent
Green functions for spin direction $\sigma$ in Nambu space via $\hat
g_\sigma=g_s\hat\tau_3+f_\sigma i\hat \tau_2$ with ubiquitous
normalization condition $g_\sigma^2+f_\sigma^2=1$. The boundary
condition for strongly spin-dependent scattering has been derived in
\cite{Eschrig:2015cea,Machon:2015wm} 
and for the present case takes the form 
\begin{equation}
   \label{eq:1}
   -i(E+\sigma \mu_BB)f_\sigma-\Delta
   g_\sigma+i\sigma f_\sigma\epsilon 
   \left\langle\frac{\sin(\phi/2)}{\cos(\phi/2)-i\sigma g_\sigma\sin(\phi/2)}\right\rangle=0\,.
\end{equation}
Here $\mu_BB$ is the Zeeman energy due to an external magnetic field,
$\Delta$ the self-consistent pair potential,
$\epsilon=r_SE_{Th}G_Q/G$ an parameter determining the effective
influence of the interface on the superconductor with the Thouless
energy $E_{Th}$, the conductance per area $G$ of the superconducting film in
perpendicular direction, the quantum conductance $G_Q=2e^2/h$ and the
fraction $r_S$ of spin-active scattering channels at the
interface. $\langle\cdots\rangle$ denotes a suitable average over the
spin-dependent interfacial phase shifts.

\begin{figure}[tb]
  \centering
  \includegraphics[width=10cm]{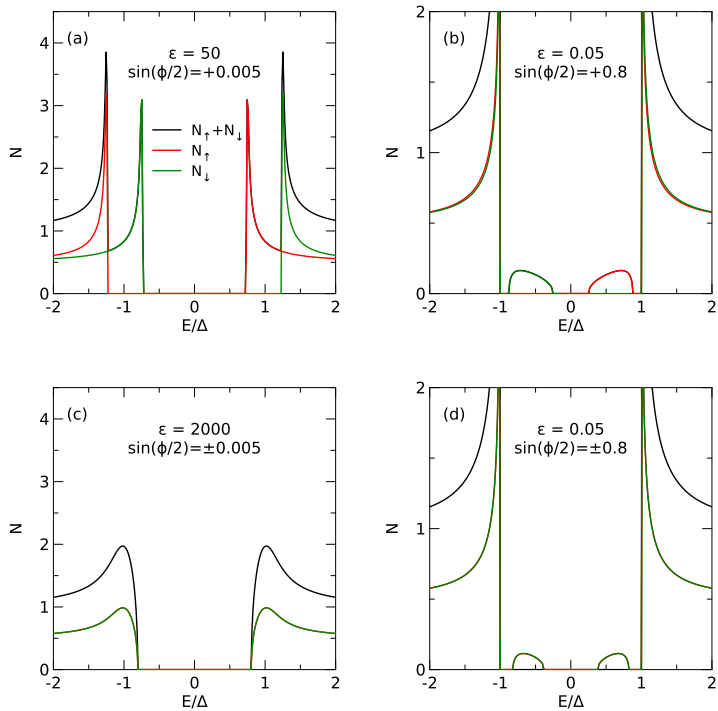}
  \caption{Spin-dependent and total density of states for four typical
    parameter sets: Small vs. large spin-mixing angles (left/right) and spin-polarized density of states with a single spin-mixing angle vs. spin averaging with two opposite spin-mixing angles (top/bottom). See text for details.}
  \label{fig:sdos}
\end{figure}

This equation is solved numerically and some results are shown in Fig.~\ref{fig:sdos}.
Figs.~\ref{fig:sdos}(a) and (b) illustrate two limiting cases with a
single spin-mixing angle. For weak spin mixing,
Fig.~\ref{fig:sdos}(a), spin-active scattering is equivalent to a
Zeeman field, as has been noted earlier by expanding the boundary
conditions in orders of $\phi$ \cite{Cottet:2009hya}. 
For strong spin mixing, Fig.~\ref{fig:sdos}(b), energy bands develop
within the energy gap of the superconductor. These bands are fully
spin polarized, in analogy to the Andreev bound state in a
single-channel superconductor-ferromagnet point contact
\cite{fogelstrom2000,zhao2004,Cottet:2008bc}.  
The subgap energy bands are remarkably similar to the well-known Shiba
bands formed by bound states at spinfull impurities
\cite{yu1965,SHIBA:1968hy,RUSINOV:1969tn}, studied experimentally in \cite{Yazdani:1997gm,Ruby:2016ec,Island:2017cc}. For ferromagnetically
aligned impurities, the Shiba bands are also predicted to be spin
polarized \cite{persson2015}. This raises the question how spin-active
scattering at interfaces in the diffusive limit is linked to
spin-dependent scattering at randomly distributed magnetic
impurities. We therefore can try to  to map the equations 
for the DOS in a superconductor in contact with a strong ferromagnetic
insulator to the known behavior of strong magnetic impurity bands
(strong means here in the presence of Shiba states). The old problem
has been treated in \cite{Zittartz:2016wq} extending the work by Shiba
\cite{yu1965,SHIBA:1968hy,RUSINOV:1969tn}. Shiba has shown that a
strong magnetic impurity in a superconductor leads to the formation of
a bound state with energy 
\begin{equation}
E_B=\Delta\frac{1-\gamma^2}{1+\gamma^2}\equiv\Delta\varepsilon_B\,
\end{equation}
where $\Delta$ is the superconducting gap energy and scattering parameter $\gamma^2=\pi^2 S(S+1) J^2 N_0^2$. Zittartz and coworkers have shown that the equation for the Green function can be cast in the form
\begin{equation}
	\label{eq:sc}
	\frac{\omega}{\Delta}=u\left(1-i\frac{\Gamma}{\Delta}\frac{\sqrt{ u^2-1}}{u^2-\varepsilon_B^2}\right)
\end{equation}
Here $\Gamma$ is a parameter related to the spin-flip scattering rate
and depends e.g. on the impurity concentration.
The density of states follows from $N(E)=\textrm{Re}[u/\sqrt{1-u^2}]$
and some resulting forms of the DOS are found in \cite{Zittartz:2016wq}

To compare this with our result using the spin-mixing angle we note that
spin mixing leads not only to pair breaking, but simultaneously adds an
exchange energy shift. To overcome this, let us assume that we have the same
number of positive and negative phase shifts. This corresponds roughly
to a random orientation of the impurity spin, which is the same
assumption as in \cite{Zittartz:2016wq}. Hence, we obtain the
following equation from \ref{eq:1} 
\begin{equation}
  0 =  i\omega f+\Delta g+\frac{\epsilon}{2} \sum_{\alpha=\pm}\frac{i\sigma \alpha s f}{c+i\sigma\alpha s g}
  =  i\omega f+\Delta g+\epsilon s^2\frac{fg}{c^2+s^2g^2}
\end{equation}
Here, we introduced $c=\cos(\delta\phi/2)$, $s=\sin(\delta\phi/2)$ and
lumped other parameters into the rate $\Gamma$. Note that the
spin-dependence signaled by $\sigma$ has dropped out, since $\hat
g_\pm$ fulfill the same equations in this case in the absence of an
external field. So we obtain spin-independent Green functions. We can
map this exactly onto the Equation (\ref{eq:sc}) by identifying
$u=ig/f$ with the result
\begin{equation}
	\frac{\omega}{\Delta}=u\left(1+i\frac{\epsilon\sin(\phi/2)^2}{\Delta}\frac{\sqrt{u^2-1}}{u^2-c^2}\right)
\end{equation}
from which we read the scattering rate of Ref.~\cite{Zittartz:2016wq} $\Gamma=\epsilon\sin(\phi/2)^2$. We have used that $g=u/\sqrt{ u^2-1}$ which follows from the normalization condition. Hence, the formulas match exactly and we can identify the Shiba bound state energy $\varepsilon_B=c=\cos(\delta\phi/2)$, which is exactly expected \cite{Hubler:2012jo} and the main result here.  To illustrate this correspondence we plot two further examples in Fig.~\ref{fig:sdos}(c) and (d), where we use two spin-mixing angles of equal weight and magnitude, but opposite sign. In this case, the density of states is spin-degenerate. For weak spin-mixing, an Abrikosov-Gor'kov type broadening of the density of states is observed. 
The identification of the boundary condition to second order in $\phi$ with
the Abrikosov-Gor'kov pair breaking mechanism has already been noted earlier \cite{Cottet:2009hya}. The effective pair-breaking rate in this case is given by $\Gamma=\epsilon\sin(\phi/2)^2$, whereas the effective Zeeman splitting in Fig.~\ref{fig:sdos}(a) is given by $\epsilon\sin(\phi/2)$. Therefore, we chose a larger $\epsilon$ for Fig.~\ref{fig:sdos}(c) to illustrate the pair-breaking effect.
Fig.~\ref{fig:sdos}(d) shows the case of strong spin-mixing, where the well-known spin-degenerate Shiba bands in the superconducting gap are recovered. 

\begin{figure}[tb]
  \centering
  \includegraphics[width=10cm]{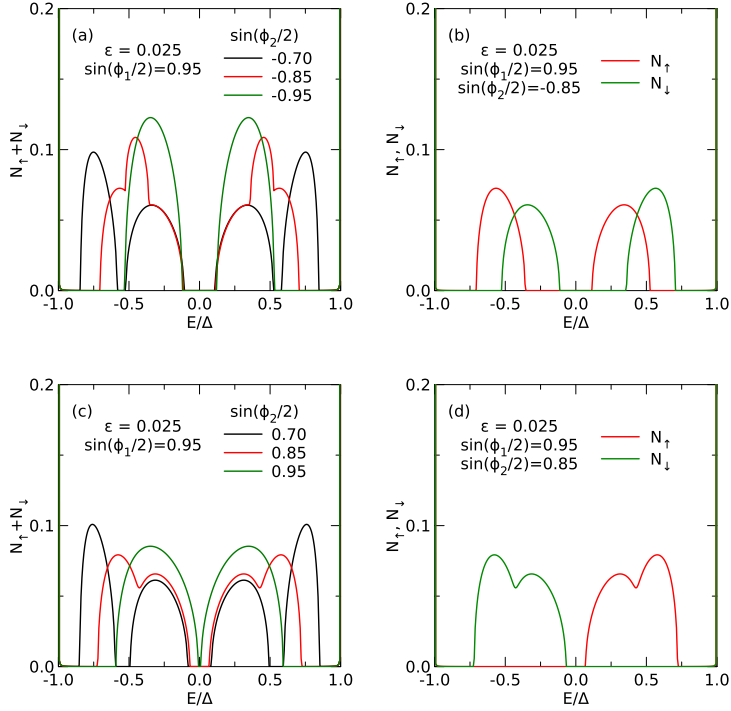}
  \caption{Effect of two interacting Shiba bands. The weights are
    taken equal and the spin-mixing parameter of one band is
    varied. (a) and (c) show the difference between aligned (same sign
    of $\phi$) and anti-aligned (opposite signs of $\phi$) Shiba
    impurity band. The strongly different behaviors can be explained
    by the spin-resolved densities of states in (b) and (d). Whereas
    the equal-spin impurity band strongly hybridized in (d), the
    oppositely polarized bands cross without interaction.}
  \label{fig:2shiba}
\end{figure}

To further explore the consequences of the strongly spin-dependent
boundary condition, we study the effect of two interacting Shiba
bands. This means we take two types of spin active channels with spin
mixing angles $\phi_1$ and $\phi_2$. We assume they are
described by the effective spin-flip rate $\epsilon$ and assume the
same number of both types of scatters. Some exemplary results are
plotted in Fig.~\ref{fig:2shiba}. In this figure we fix the spin
mixing angle of the first band $\sin\phi_1/2=0.85$ and vary the second
spin mixing angle. Fig.~\ref{fig:2shiba} (a) and (c) show the total density of
states for different spin mixing angle with the opposite (a) and the
same (c) sign. For the case with $\sin\phi_2/2=-0.7$ (black curve), the two Shiba
bands are not overlapping in energy and, hence, the total density of
states is simply the sum of the two Shiba bands. Note, that the
two bands with positive energy have opposite spin polarizations. For
the case of opposite spin mixing angles $\sin\phi_2/2=-\sin\phi_1/2$
(green curve), the bands are at the same energy and, according to the
argument before, the total density of states is unpolarized. For the
case of almost similar magnitudes of the spin mixing angles
$\sin\phi_2/2=-0.85$, the total density of states has a nontrivial
shape with a central peak emerging in a broad background. This
behavior can be explained by looking at Fig.~\ref{fig:2shiba} (b)
showing the spin-resolved density of states, which do not interact in
the overlap regime, giving ride to the peculiar peaked behavior of the
total density of states in this case. In Fig.~\ref{fig:2shiba} (b) we
show the case of two similar spin mixing angles of similar size and
the same sign. Obviously nothing prevents the two equally polarized Shiba bands to
interact at the same energy, which results in a simply hybridization
in the case of almost equal spin mixing angles (red curve in
Fig.~\ref{fig:2shiba} (c)). In this case the Shiba bands remain
spin-polarized, as is illustrated in the  Fig.~\ref{fig:2shiba}
(d). Finally, we note that the case of equal of the spin
mixing angle  (green curve in Fig.~\ref{fig:2shiba} (c)) differs
quantitatively from the case case of opposite signs (green curve in
Fig.~\ref{fig:2shiba} (a)).  

In conclusion, we have show that the strongly spin-dependent
scattering at an interface to a  magnetic insulator has a similar
effect as scattering of spinfull impurities, which lead to the
formation of Yu-Shiba-Rusinov states.

We acknowledge discussions with P. Machon. This work was financially
supported by the DFG through SPP 1538 \textit{Spincaloric Transport} and Grant No. BE-4422/2-1.


\bibliography{FIS_Shiba}

\end{document}